\documentclass[aps,pre,twocolumn,superscriptaddress,showpacs]{revtex4}
\def\av#1{\left\langle#1\right\rangle}

\usepackage{amstext,amssymb,amsfonts,amsmath,eucal,graphicx,bm}
\usepackage{epsfig}

\begin{document}

\title{Partition of Networks into Basins of Attraction}

\author{Shai Carmi}
\affiliation{Center for Polymer Studies, Boston University, Boston,
MA 02215, USA} \affiliation{Minerva Center \& Department of Physics,
Bar-Ilan University, Ramat Gan 52900, Israel}
\author{P. L. Krapivsky}
\affiliation{Department of Physics, Boston University, Boston, MA
02215, USA} \affiliation{Theoretical Division and Center for
Nonlinear Studies, Los Alamos National Laboratory, Los Alamos, New
Mexico 87545, USA}
\author{Daniel ben-Avraham}
\affiliation{Department of Physics, Clarkson University, Potsdam NY
13699-5820, USA}

\begin{abstract}
We study partition of networks into basins of attraction based on a
steepest ascent search for the node of highest degree. Each node is
associated with, or ``attracted" to its neighbor of maximal degree,
as long as the degree is increasing. A node that has no neighbors of
higher degree is a peak, attracting all the nodes in its basin.
Maximally random scale-free networks exhibit different behavior
based on their degree distribution exponent $\gamma$: for small
$\gamma$ (broad distribution) networks are dominated by a giant
basin, whereas for large $\gamma$ (narrow distribution) there are
numerous basins, with peaks attracting mainly their nearest
neighbors. We derive expressions for the first two moments of the
number of basins. We also obtain the complete distribution of basin
sizes for a class of hierarchical deterministic scale-free networks
that resemble random nets. Finally, we generalize the problem to
regular networks and lattices where all degrees are equal, and thus
the attractiveness of a node must be determined by an assigned
weight, rather than the degree. We derive the complete distribution
of basins of attraction resulting from randomly assigned weights in
one-dimensional chains.
\end{abstract}
\date{\today}


\maketitle

\section{Introduction}

\label{intro}

Networks often have heterogenous structure, with different nodes
highly varying in their connectivity and in their roles
\cite{BA_review,Newman_review,PV_book,DM_book}. The problem of
identifying these roles and assigning nodes to communities or
modules based on their function is of great interest, with many
methods and algorithms recently proposed
\cite{community1,community2,community3,community4,community5,community6,community7,community8,community9,community10,community11,community12,community13}.
These methods aim to incorporate knowledge of the global network's
structure with information of the nodes' local connections, to
generate a network partition. However, on a more fundamental level,
nodes can be simply distinguished according to the node that is
their ``authority'', ``attractor'', or, in heterogeneous networks,
their ``hub''.

\begin{figure}
\includegraphics[width=6cm]{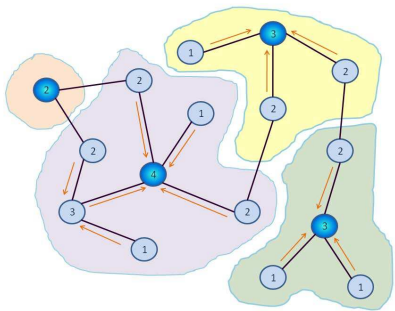}
\caption{(Color online) Decomposition of a network into basins of
attraction. The number inside each node indicates its degree, and
the arrows point from a node to its attractor --- the neighbor with
highest degree. Nodes with no neighbors of higher degree are
\emph{peaks} and are highlighted in the schematic. A peak represents
a basin of attraction: all nodes which are attracted to it belong to
its basin of attraction. The basins in this schematic have different
background colors.} \label{Schematic}
\end{figure}

The hub that each node belongs to is found by moving recursively
onto the neighbor of highest degree, or number of connections, until
the hub is reached  --- a node whose degree is greater than that of
all of its neighbors. Classifying nodes by their hubs leads to a
natural partition of the network into basins of attraction. See
Fig.~\ref{Schematic} for a schematic illustration. This partitioning
provides a quick and easy way to classify nodes based on their
relation with the network's major players, without resorting to
external information.

In general, when each node is associated with a value of a scalar
field, a ``gradient network'' emerges by replacing all the links
that emanate from a node by a single directed link that points to
the node's neighbor with the highest value of the field
\cite{gradient1,gradient2}. Thus, recursively following nodes of
highest degree is equivalent to traversing the ``gradient network''
formed by considering the scalar field defined by the degrees of the
nodes. Many properties of gradient networks have been studied, such
as the emerging degree distribution and its relation to the original
network topology, and the possibility of congestion when too few
nodes are receiving the flow that is generated by the gradient
\cite{gradient1,gradient2}. Gradient networks have also proved
useful in the analysis of energy landscapes \cite{gradient_protein},
and as the basis for new and improved synchronization
\cite{gradient_sync} and routing \cite{gradient_routing} methods.
Here, we focus on the specific case where the value associated with
each node is the degree, and thus does not require any external
information but the bare topology. The walk up the degree gradient
identifies each node with one of the network hubs.

The decomposition into steepest-ascent basins is of interest in many
systems. For example, suggested routing schemes in communication
networks involve transmitting all packets through the hub nearest to
the source \cite{routing,Huberman}. The size of the basins delimits
the performance of such routing algorithm. In a different field, an
analysis of the energy landscape's network of atomic clusters shows
that the energy of a configuration, or a node, decreases with the
number of configurations kinetically connected to it, which is its
degree \cite{Doye}. Thus, as the system is cooled and its energy
decreases, configurations with higher degrees tend to be visited.
The actual partitioning into basins determines roughly whether the
system would inevitably end up in the ``ideal glass state'' or
arrive at one of many meta-stable states, depending on the initial
conditions \cite{SupercooledReview,Sastry}.

The topology of the basins is also important if one is interested in
a local strategy for finding the most connected node. A network with
a single basin would make a steepest ascent search (in the ``degree
space") successful, while a more complicated topology would require
a more sophisticated approach. Finally, the properties of the basins
of attraction can be used to classify networks with similar degree
distributions but otherwise different topology and function.

The algorithmic aspects of the partition method are relatively
simple and will be discussed briefly below. Our main goal is to
study, analytically and numerically, the statistical properties of
the basins of attraction in ensembles of maximally random scale-free
(SF) networks \cite{BA_review,Newman_review,PV_book,DM_book}. We
find that the topology of basins (i.e. their number, sizes, hubs'
degree, etc.) shows a strong dependence on the degree distribution,
and we quantify this behavior. We then study the basins' topology in
a class of deterministic hierarchical SF networks
\cite{flowers1,flowers2} and show that it reflects some prominent
properties of the random networks. Finally, we generalize the
problem to the case where the ``attractiveness'' of each node is
determined by a random number, a `height' rather than its degree,
and derive analytical results for the basins of attraction in
regular one- and two-dimensional lattices.

\section{Definitions}
\label{DefSect}

We focus on (undirected) SF networks, i.e., networks in which node
degree is broadly distributed, usually in the form ${\cal P}(k) \sim
k^{-\gamma}~~(m \leq k \leq N)$, where $k$ is the degree, $m$ is the
minimum possible degree, $N$ is the total number of nodes, and
$\gamma>2$ is the degree exponent. Many real world networks were
shown to be scale-free with $\gamma<3$
\cite{BA_review,Newman_review,PV_book,DM_book}. We nevertheless
study networks with $\gamma \in [2,5]$, so as to reach the
near-homogenous limit where the degree is narrowly distributed. Our
networks are static and maximally random, generated according to the
configuration model \cite{MR}: we first draw nodes' degrees based on
the prescribed distribution, then randomly connect open links until
all nodes have all of their links connected.

A precise definition of basins of attraction requires dealing with
several ambiguities (e.g., how to resolve tie-breaks). We opt for
the following rules:
\begin{enumerate}
\item Start the search from node $i$ with degree $k_i$
and neighbors $j_1,j_2,...,j_{k_i}$.
\item Denote the neighbor that has the highest degree as $j_\textrm{max}$,
with degree
$k_{j_\textrm{max}}=\max\{k_{j_1},k_{j_2},...,k_{j_{k_i}}\}$. If the
highest degree is shared by more than one neighbor, choose one of
them arbitrarily.
\item If $k_i<k_{j_\textrm{max}}$, $i$ is attracted to $j_\textrm{max}$ and both belong to
the same basin of attraction.
\item If $k_i \geq k_{j_\textrm{max}}$, node $i$ is a \emph{peak}, and is attracted to
itself, forming a basin of attraction with all nodes (if exist) that
are attracted to it.
\item Repeat for all unassigned nodes as the root of the search. Each node
now belongs to exactly one basin of attraction.
\end{enumerate}

Note that we require $k_i$ to be strictly smaller than
$k_{j_\textrm{max}}$ for $i$ to be attracted to $j_\textrm{max}$; in
other words, a node may be a peak even if it has neighbors with
equal degree. This choice saves us from delving into further
subtleties. The results are qualitatively the same independent on
details of the definition (see Appendix \ref{alternative} for a
short discussion).

A simple and fast partitioning algorithm relies on scanning the
nodes in descending degree order. Then, each node is either
designated as a peak, or assigned to the basin of its neighbor with
highest degree. Because we scan by degree order, we are guaranteed
that the neighbor was already assigned to a basin. Thus, the running
time of the algorithm (for sparse networks) is of the order of
$N\log{N}$, the time it takes to sort the nodes \cite{Coreman}.

\section{Topology of basins of attraction in random scale-free networks}
\label{simulations}

We attempt to capture the topology of the basins through a few
representative quantities which we define below. Denote the total
number of basins by $N_b$. Define the density of basins as the
number of basins \emph{per node} and denote it as $n_b\equiv N_b/N$.
Denote next the basin size by $s$. The probability of a basin to be
of size $s$ is $P(s)$. A related quantity is the probability of a
node to belong to a basin of size $s$: $Q(s)=N_bP(s)s/N=n_bsP(s)$
(and a particularly interesting case is the probability of a node to
be a solitary basin $Q(1)$). Both $P$ and $Q$ are normalized
probability distributions. Other measures of interest are the
degrees of the peaks and the size of the largest basin $S$.

\begin{figure}
\includegraphics[width=8cm]{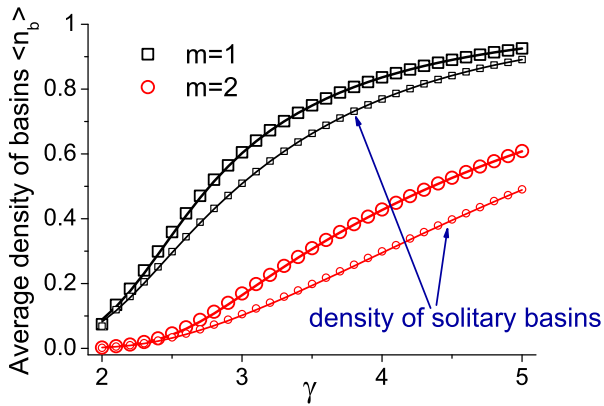}
\caption{Average density of basins in SF networks.  Plotted is
$\av{n_b}$ vs. $\gamma$, as well as the density of solitary basins,
$Q(1)$. Simulation results (symbols), for networks with $N=1000$ and
$m=1,2$, are matched by theory, Eqs.~(\ref{p_peak}) and
(\ref{q1_eq}) (solid lines).} \label{NumBasins}
\end{figure}

\begin{figure}
\includegraphics[width=8cm]{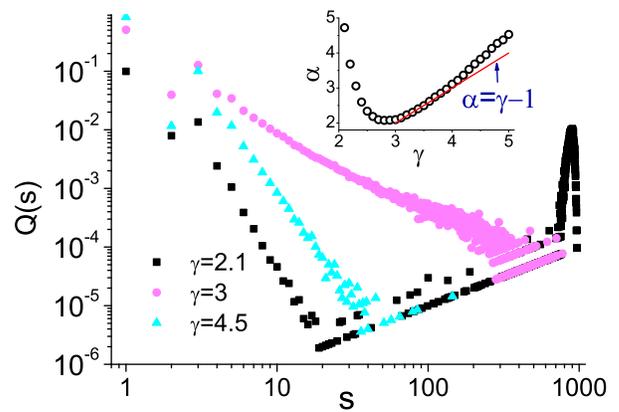}
\caption{Distribution of basin sizes in SF networks. Plotted is
$Q(s)$ for three values of $\gamma$ ($N=1000$, $m=1$). Inset: The
basin exponent $\alpha$, characterizing the power-law decay at small
$s$, plotted vs. the degree exponent $\gamma$.} \label{Qofs}
\end{figure}

\begin{figure}
\includegraphics[width=8cm]{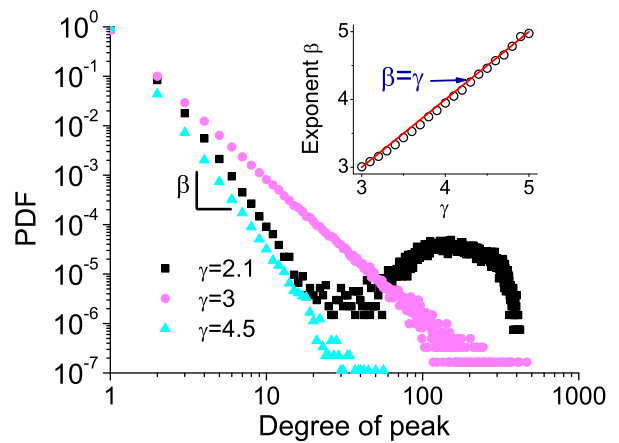}
\caption{Degree of the peaks in SF networks. Plotted is the
distribution of the degree of the peaks for three values of $\gamma$
($N=1000$ and $m=1$). Inset: For large $\gamma$, the exponent
$\beta$ characterizing the decay of the pdf (circles) follows
$\gamma$ (solid line) very closely. } \label{DegreePeak}
\end{figure}

\begin{figure}
\includegraphics[width=8cm]{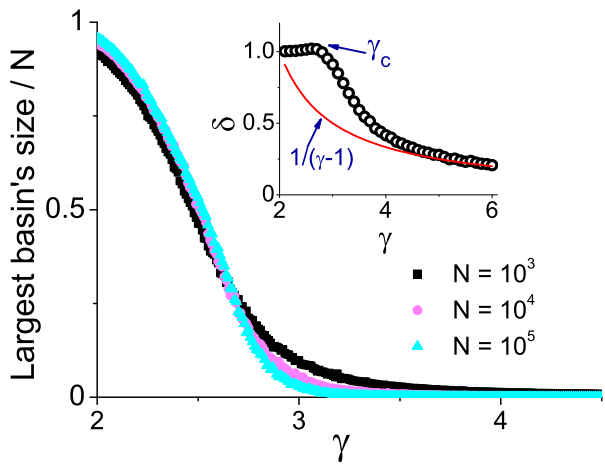}
\caption{Size of the largest basin. Plotted is the relative size of
the largest basin ($S/N$) vs. $\gamma$ for SF networks with
$N=10^3$, $10^4$, and $10^5$, and $m=1$. Inset: The largest basin's
exponent $\delta$ ($S \sim N^{\delta}$). For $\gamma \lesssim
\gamma_c$, $\delta \simeq 1$ ($S \sim N$ is a giant basin), while
for large $\gamma$, $\delta$ approaches $1/(\gamma-1)$.}
\label{MaxBasin}
\end{figure}

In Figs.~\ref{NumBasins}-\ref{MaxBasin} we present simulation
results for SF networks with $N=1000$, minimum degree $m=1,2$ and
varying $\gamma$. The following picture emerges from the results.
For small $\gamma$ close to 2, the network is dominated by one hub,
attracting most of the nodes to form a giant basin. Thus, the number
of basins is relatively small and the size of the largest basin is
narrowly distributed about $S \sim N$. The sizes of the basins and
the degrees of the peaks show a bimodal distribution: a peak close
to $N$ and a fast decay for small basins which are not included in
the giant basin.

For large $\gamma$, a different behavior is observed. The number of
basins, $N_b$, is large, and most of the basins are small. The
largest basin is no longer giant, and its average size scales as $S
\sim N^{\delta}~(\delta<1)$. The distribution of the degrees of the
peaks approaches the degree distribution of the entire network. The
distribution of basin sizes now exhibits power-law scaling for small
$s$: $Q(s)\sim s^{-\alpha}$ (or $P(s)\sim s^{-(\alpha+1)}$). We term
$\alpha$ the \emph{basin exponent}. The minimal degree $m$
significantly influences the basins count. For $m=1$ the network is
usually fragmented, and thus many basins can form. For $m \geq 2$
the network is connected and consists of a single component, so the
number of basins is smaller.

The crossover between the two limiting cases of networks with a
giant basin and networks fragmented to many basins is at about
$\gamma_c \approx 2.8$. This is revealed by the behavior of the size
of the largest basin $S$: while $S\sim N$ for $\gamma<\gamma_c$,
there is no longer a giant basin for $\gamma>\gamma_c$ and $S \sim
N^{\delta}$ with $\delta<1$ (Fig. \ref{MaxBasin}). The minimum in
$\alpha$ also occurs at $\gamma \approx 2.8$ (Fig. \ref{Qofs}), and
we hypothesize that it is another reflection of the transition.

\section{Theory}
\label{Theory}

\subsection{Random scale-free networks}

\subsubsection{The giant basin}

The transition between a network with giant basin to a network
fragmented to many basins is observed in the simulations at about
$\gamma\approx 2.8$. Interesting questions are whether this
transition becomes sharp for infinite systems, and what is the value
of $\gamma_c$ for $N \rightarrow \infty$.

A simple argument suggests that for infinite networks a sharp
transition occurs at $\gamma_c=3$. To understand that, consider
first the probability of a given node $i$ of degree $k$ to be a
peak,
\begin{equation}
\label{p_peak_k} \Pr\{\mbox{\emph{i} is a peak}|k_i=k\} = \left[
\sum_{k'=m}^k \frac{k'{\cal P}(k')}{\av{k}}\right]^k,
\end{equation}
where ${\cal P}(k)$ is the degree distribution, $\av{k}$ is the
average degree, and $\frac{k'{\cal P}(k')}{\av{k}}$ is the
probability that a neighboring node (which is a node followed by a
random link) has degree $k'$ \cite{intentional}. Equation
(\ref{p_peak_k}) results from the requirement that none of $i$'s $k$
neighbors have degree higher than $k$. For large $k$, we substitute
${\cal P}(k)\simeq A\,k^{-\gamma}$ and approximate the sum as an
integral
\begin{eqnarray*}
\sum_{k'=m}^k k'{\cal P}(k') &=& \av{k} - \sum_{k'=k+1}^\infty k'{\cal P}(k') \\
&\simeq & \av{k} - \frac{A}{\gamma-2}\,k^{2-\gamma}
\end{eqnarray*}
For $k\gg1$, Eq.~\eqref{p_peak_k} becomes ($\gamma>2$)
\begin{equation*}
\Pr\{\mbox{\emph{i} is a peak}|k_i=k\} =
\left[1-\frac{B}{k^{\gamma-2}}\right]^k \approx \exp\left(-B
k^{3-\gamma}\right),
\end{equation*}
where $B = A/[\av{k}(\gamma-2)]$. Thus, for $\gamma<3$ the
probability of a node to be a peak is small, and approaches zero for
large $k$. Therefore, only the node with the largest degree in the
network can be a peak, and it will attract the giant basin. For
$\gamma>3$, every node with large degree is almost surely a peak.
For even larger $\gamma$, $\gamma>\gamma^*>3$ (where $\gamma^*$ is
determined by the small $k$ properties of ${\cal P}(k)$) there is no
longer a giant component in the network. In that case, the size of
the largest component scales as $N^{1/(\gamma-1)}$ \cite{Largest}.
The maximal degree of the network has the same scaling. Since the
size of largest basin is at least the maximal degree, but cannot
exceed the size of the largest component, we conclude that for
$\gamma>\gamma^*>3$, $S \sim N^{\delta}$ with $\delta=1/(\gamma-1)$.
Simulation results support this scaling, (inset of Fig.
\ref{MaxBasin}), but it is not known whether a transition is
expected, for infinite network, at $\gamma^*$.

Another heuristic argument in favor of the phase transition at
$\gamma_c=3$ is the following. Consider two nodes with degrees $k_1$
and $k_2$ close to the maximal degree $K\sim N^{1/(\gamma-1)}$
\cite{resilience}. The probability that these nodes are connected is
proportional to $k_1k_2/N$ \cite{agata}. Thus, the probability of
the two hubs to be connected scales as $N^{(3-\gamma)/(\gamma-1)}$.
Hence, for $\gamma<3$ the two largest hubs are almost surely
connected. The hub with the larger degree attracts the smaller hub,
together with its entire basin, to form the giant basin. These
arguments are supported by simulation results for increasing values
of $N$ (Fig. \ref{MaxBasin}).

\subsubsection{Number of basins and basin sizes}
\label{fourb}

While we could not obtain a complete derivation of $P(s)$ or $Q(s)$
for random static scale-free networks, it is possible to obtain
analytic results for a few chief quantities. Below we derive an
exact expression for $\av{n_b}$ as well as reasonable approximations
for $\mbox{Var}(n_b)$ and $Q(1)$.

Clearly, the number of peaks is equal to the number of basins. Thus,
the average basin concentration $\av{n_b}=\av{N_b}/N$ is equal to
the probability of a node to be a peak. The probability of a given
node $i$ with degree $k$ to be a peak is given in Eq.
(\ref{p_peak_k}). If the degree of $i$ is not specified, we must
condition over all possible degrees. Thus,
\begin{equation}
\label{p_peak}
\av{n_b} = \sum_{k=m}^{\infty}{\cal P}(k)\left[
\sum_{k'=m}^k \frac{k'{\cal P}(k')}{\av{k}}\right]^k
\end{equation}
Plugging the degree distribution into \eqref{p_peak} completes the
derivation; e.g., for SF networks we substitute ${\cal
P}(k)=k^{-\gamma}/\sum_{k'=m}^{\infty}k'^{-\gamma}$.
A comparison of
Eq.~(\ref{p_peak}) with simulations yields a perfect agreement
(Fig.~\ref{NumBasins}).

Many real-life networks \cite{log}, and in particular growing ones,
have $\gamma$ close to 2 and accordingly, a logarithmically
diverging average degree $\av{k}\sim \ln N$. Consequently, the $k=m$
term dominates Eq. (\ref{p_peak}):
\begin{equation}
\label{gamma2} \av{n_b} \sim {\cal P}(m)\left[m{\cal P}(m)\right]^m
(\ln N)^{-m} +\mathcal{O}[(\ln N)^{-(m+1)}],
\end{equation}
and we expect $\av{n_b} \sim [\log{N}]^{-m} \rightarrow 0$ for $N
\rightarrow \infty$ (such that the finite value of $\av{n_b}$ at
$\gamma \rightarrow 2$ for $m=1$ in Fig. \ref{NumBasins} is a finite
size effect).

The calculation of the variance of $N_b$ is more involved since the
joint probabilities for multiple peaks are not independent. An
approximate expression is given in Appendix \ref{theory_app} and is
plotted in Fig.~\ref{Variance_fig}. Rather than the full
distribution $Q(s)$, we focus on solitary basins (of size $s=1$),
which account for the bulk of basins. In Appendix \ref{theory_app}
we derive an approximation for $Q(1)$ which is extremely close to
simulation results (see Fig.~\ref{NumBasins}).

\subsection{Hierarchical networks}
\label{flowers}

Deterministic hierarchical scale-free networks provide a unique
opportunity for an analytical treatment of networks with broad
degree distribution~\cite{flowers1,flowers2}. In the following we
derive analytical results for the basins topology, which reproduce
to some extent the results for random SF networks. In particular,
hierarchical networks have a giant basin for small $\gamma$, and a
power-law distribution of the basin sizes $P(s)$ for large $\gamma$,
just as was found in Section \ref{simulations} for the random
networks.

Hierarchical scale-free networks~\cite{flowers1,flowers2} are
construc\-ted in a recursive fashion: in $(u,v)$-flowers, each link
in generation $n$ is replaced by two parallel paths consisting of
$u$ and $v$ links, to yield generation $n+1$~(Fig.~\ref{flowers});
and in $(u,v)$-trees, defined in analogy to the flowers, we obtain
generation $n+1$ of a $(u,v)$-tree by replacing every link in
generation $n$ with a chain of $u$ links, and attaching to each of
its endpoints chains of $v/2$ links (assuming $v$ is even). A
natural choice for the genus of flowers in generation $n=1$ is a
cycle graph (a ring) consisting of $u+v\equiv w$ links and nodes.
$(u,v)$-flowers and trees were shown to have degree distribution of
the form ${\cal P}(k) \sim k^{-\gamma}$, with
$\gamma=1+\frac{\ln{w}}{\ln2}$, and are thus scale-free. Considering
shortest paths, $(u,v)$-nets with $u=1$ are small-worlds and are
otherwise fractals \cite{flowers1,flowers2}. These and other
topological properties, such as clustering and degree-degree
correlations make them suitable models for real-life complex
networks~\cite{BA_review,Newman_review,song,web,assortative}.

\begin{figure}
\includegraphics[width=8cm]{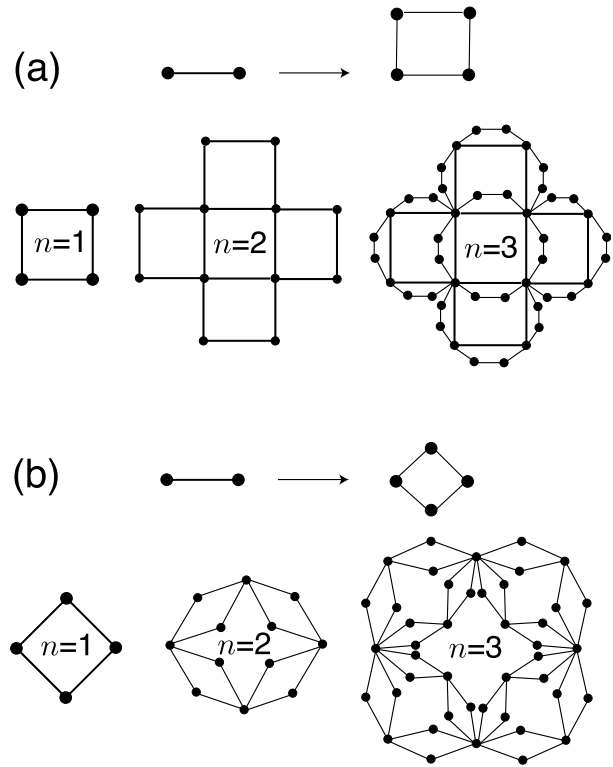}
\caption{Hierarcical scale-free ($u,v$)-flowers. Shown are two
examples of networks with degree exponent $\gamma=\ln(u+v)/\ln2=3$,
with (a)~$u=1$, $v=3$, and (b)~$u=2$, $v=2$.  In both cases the top
of the figure illustrates the edge replacement scheme, to two
parallel paths of $u$ and $v$ edges, while the bottom of the figure
shows flowers obtained in this way to generations $n=1$, 2, and 3.}
\label{flowers}
\end{figure}

We have derived the complete distribution of $P(s)$ for all
$(u,v)$-flowers and trees. This is a tedious exercise in real-space
renormalization (verified numerically on a computer)  that adds
little physical insight. We thus limit the discussion to the results
themselves.

For $(1,2)$ and $(1,3)$-flowers, and $(1,2)$-trees, all nodes are
evenly split between the $w$ basins peaked in the nodes forming the
$n=1$ generation. Thus, this case corresponds to the small $\gamma$
limit of random SF networks, where a giant basin attracts all nodes.

In $(1,v)$-flowers with $v\geq 4$ (which corresponds to $\gamma>3$),
basins of size $b_m=\frac{2}{3}4^m+\frac{1}{3}$ ($m=0,1,...,n-2$)
appear $(v-3)w^{n-m-1}$ times. Thus, basins of size $s\sim4^m$ occur
with frequency $s^{-\ln w/\ln4}$. Because the possible basin sizes
are not continuous but are exponentially spaced, this leads to a
power-law distribution $P(s) \sim s^{-(\alpha+1)}$, with basin
exponent $\alpha=\frac{\ln{w}}{\ln4}=(\gamma-1)/2$.

In $(1,v)$-trees the situation is qualitatively similar, but more
subtle, with different results for $v=4$ and $v>4$. For
$(1,4)$-trees, we find that basins of size $b_m$ ($m=2,3,...,n-1$)
appear $2w^{n-m-1}$ times. Here $b_m \sim Ar^m$ ($m \gg 1$) where
$r$ is the larger root of $r^2-5r+2=0$, or
$r=\frac{5+\sqrt{17}}{2}=4.56$. Thus, $P(s)\sim s^{-(\alpha+1)}$
with $\alpha=\frac{\log{5}}{\log{r}}=1.06$. For $(1,v)$-trees with
$v>4$, basins of size $b_m=\frac{5}{24}4^m+\frac{2}{3}$
($m=2,3,...,n-1$) appear $2w^{n-m}$ times, so $P(s) \sim
s^{-(\alpha+1)}$ with $\alpha=\frac{\ln{w}}{\ln{4}}$, as in
$(1,v)$-flowers. The size of the largest basin for all $(1,v)$-nets
(except $(1,4)$-trees) is $S \sim N^{\delta}$ with
$\delta=2/(\gamma-1)$.

In $(u,v)$-nets with $u \geq 2$ and $v \geq 2$ the number of basins
of size $2^m+1$ ($1.5\cdot2^m+1$ for $(u,4)$-trees),
$m=3,4,...,n-1$, is $(w-2)w^{n-m}$ for flowers and
$[w-3+2/w]w^{n-m}$ for trees. This simply leads to $P(s)\sim
s^{-\gamma}$, or $\alpha=\gamma-1$. Essentially, due to the strong
disassortative nature of $(u,v)$-nets with $u \geq 2$ and $v \geq
2$, the basins in these networks typically consist of a peak and its
immediate neighbors, so the basins sizes mirror the degree
distribution. Indeed, the size of the largest basin $S\sim
N^{\frac{1}{\gamma-1}}$, has the same scaling as the largest degree
\cite{flowers1,resilience}.

To summarize, with the exception of $(1,4)$-trees:
\begin{eqnarray}
\alpha_{(1,v)\textrm{-nets}} &=& \left\{ \begin{array}{ll}
(\gamma-1)/2 & \gamma>3, \\
w \textrm{ giant basins} & \gamma \leq 3.\\
\end{array} \right. \\
\alpha_{(2,v)\textrm{-nets}} &=& \gamma-1,\qquad ~~~ \gamma \geq 3\,.
\end{eqnarray}
Key features revealed by this analysis compare favorably with the
results in random scale-free networks. The giant basins found for
hierarchical nets with $\gamma\leq3$ parallels the low-$\gamma$
phase found in random nets. The power-law decay found for $\gamma>3$
agrees with the findings for large $\gamma$ in random nets, as does
the increase of $\alpha$ with increasing~$\gamma$.

\section{Random surfaces}

The decomposition of a network into degree-based basins of
attraction is a special case of a general problem of finding the
basins when the attractiveness of a node is determined by a certain
attribute. The association of a scalar field with the network nodes
and the emergence of a ``gradient network'' were suggested in
\cite{gradient1,gradient2} and discussed in Section \ref{intro}.
Here, our main interest is in the basins of attractions induced by
the external field. In particular, determining the attractiveness of
a node by an external parameter allows the basins of attraction to
be defined in \emph{regular} networks or lattices where all sites
have the same degree. As a basic example, we discuss one- and
two-dimensional lattices where each node is assigned a random
\emph{height} (or potential energy, density, etc.) The understanding
of the topology of such random surfaces is of much importance
\cite{RandomLandscape,fugacity}. For example, the number of peaks
determines the number of possible non-satisfied bonds in a spin
glass \cite{SpinNetwork,Derrida} or the ``roof'' of the surface in
ballistic growth models \cite{Oshanin}.

\begin{figure}[htb]
\centering
\includegraphics[width=7cm]{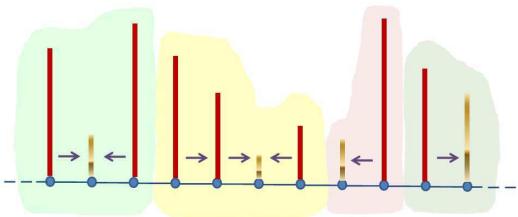} \vspace{-1.5cm}
\caption{\small {(Color online) Schematic representation of a random
surface in one dimension. Each node is attracted to its shortest
neighbor (with periodic boundary conditions). The four highlighted
nodes are valleys
--- their neighbors are taller. All nodes attracted to a valley
belong to its basin of attraction. The four basins in the drawing
are shown in different background colors.}} \label{lattice}
\end{figure}

The height $h_i$ of lattice site $i$ is taken from some distribution
(independently of the other lattice sites). Without loss of
generality, one may assume the distribution is uniform, in the
interval $[0,1]$. Nodes are attracted to their \emph{shortest}
neighbor, so that the surface is energy-like (Fig.~\ref{lattice}).
The topology of the basins, in this case, has a clear physical
interpretation: Put a particle in each node of the lattice and let
the particles follow paths of steepest descent. When the system
stops evolving, the number of particles $s$ in each minimum is the
size of its basin of attraction.

In one dimension, each point on the surface is either a local
maximum (peak), a local minimum (valley), or it has one taller and one shorter
neighbor. To find the density of peaks we look at any
three consecutive heights $(h_1,h_2,h_3)$ and notice that the
probability that $h_2$ is maximal is $1/3$. Similarly, the density of
valleys is also $1/3$. The variance in the number of peaks/valleys can be
derived following similar steps as for networks (Appendix
\ref{theory_app}) and turns out to be $2N/45$ \cite{SpinNetwork}.

Let us calculate the probability of a node to be a valley of a basin
of size $s$, $R(s)=\av{n_b}P(s)$. The minimal size $s=1$ of the
basin is obtained in the situation when the minimum is surrounded by
two taller heights whose other adjacent heights are shorter than the
minimum. If $h$ is the height of the minimum, the above situation
occurs with probability $[h(1-h)]^2$. Integrating over $h$ we find
the density of smallest basins
\begin{equation}
\label{R1} R(1) = \int_0^1 dh\,h^2(1-h)^2 = \frac{1}{30}\,.
\end{equation}

For $s\geq 2$, the density of basins of attraction of size $s$ is
given by
\begin{equation}
\label{Rs} R(s) = \frac{2^{s+3}}{(s+4)!}\,s(s+3) -
\frac{4(s^2+3s+1)}{(s+3)!}\,.
\end{equation}
The derivation of this result is presented in Appendix \ref{der}.
One can verify the validity of both the normalization requirement
and the density of valleys:
\begin{equation*}
\sum_{s\geq 1} sR(s)=1\,, \quad \sum_{s\geq 1}
R(s)=\frac{1}{3}=\av{n_b}\,.
\end{equation*}
For large $s$, $R(s) \sim 1/s!$, which decays much faster than the
power-law decay observed for networks.

In two dimensions, basins of attraction are similarly defined as the
set of all nodes which are attracted to a given valley. We limit
ourselves to the analytical computation of $R(s=1)$, as larger
basins of attraction seem to require very tedious calculations. Let
$h$ be the height of the minimum of an $s=1$-basin of attraction.
The adjacent four heights must be taller, which happens with
probability $(1-h)^4$. We write
\begin{equation}
\label{R1:int} R(1) = \int_0^1 dh\,(1-h)^4 \sigma(h)\,,
\end{equation}
and the chief problem is to determine the probability $\sigma(h)$
that for each of the 4 adjacent sites there is a neighbor which is
shorter than $h$. Let $\{x_i\}$ be the heights of  diagonal sites
$(\pm 1,\pm 1)$, and $\{y_j\}$ the heights of the sites $(0,\pm 2)$
and $(\pm 2, 0)$  [we set the minimum at the origin]. The probability $\sigma(h)$ is given by
\begin{eqnarray*}
\sigma(h) &=& (1-h)^4 h^4 + 4h(1-h)^3 h^2\\
&+& 2h^2(1-h)^2 + 4h^2(1-h)^2 h\\
&+& 4h^3(1-h) + h^4\,.
\end{eqnarray*}
Indeed, one possibility is that all the $x_i$ exceed $h$, and then all
the $y_j$ must be shorter than $h$. This happens with probability
$(1-h)^4 h^4$. If exactly three of the $x_i$ are taller than $h$, there
should be exactly two $y_j$ that are shorter than $h$. This explains
the term $4h(1-h)^3 h^2$. For  the case that two of the $x_i$ are taller
 and two shorter than $h$, consideration of their exact locations
 leads to the term $2h^2(1-h)^2 + 4h^2(1-h)^2
h$. Finally, when at most one $x_i$ is taller than $h$, there is no requirement on the
$y_j$. Performing the integral in
\eqref{R1:int} we obtain
\begin{equation}
\label{R1:2D} R(1) = \frac{109}{4290}\,.
\end{equation}

In the infinite dimensional case, the random surface is defined on
top of a network, as in gradient networks
\cite{gradient1,gradient2}. The only quantity that seems easily
calculable is the average number of basins $\av{n_b}$: the
probability of a node of degree $k$ to be a valley, for randomly
distributed heights, is simply $1/(k+1)$. Thus, for a network,
\begin{equation}
\label{n_bN} \av{n_b} = \sum_k{\cal P}(k)/(k+1)\,.
\end{equation}

\section{Summary and Discussion}

In summary, we have introduced a process of steepest ascent that
partitions complex networks into basins of attraction --- subsets of
nodes that are attracted to the same peak, the node of highest
degree in the basin. For random scale-free networks we find a
transition between networks dominated by a giant basin comprising
the majority of the nodes, for $\gamma\lesssim \gamma_c$, to
numerous, fragmented basins, for $\gamma\gtrsim \gamma_c$. We find
numerically that $\gamma_c\approx 2.8$, while theoretical arguments
indicate that for $N \rightarrow \infty$, $\gamma_c=3$. Both above
and below the transition point, the distribution of finite basins
has a power-law tail $s^{-(\alpha+1)}$, where $\alpha$, the basin
exponent, exhibits a non-trivial dependence upon the degree exponent
$\gamma$. An exact analysis of deterministic hierarchical scale-free
nets exhibits some of these features.

A comprehensive description of the complete distribution of basins
sizes for static random scale-free networks remains a challenge.
Furthermore, other types of networks might exhibit a different basin
topology. In particular, randomly growing networks
\cite{BA_review,KR}, Erd\H os-R\'enyi networks \cite{Bollobas}, and
networks with correlations (for example, degree-degree correlations)
are of interest and are left for future study.

In a sense, associating each node with a hub and the identification
of basins of attraction provides a partition of the network into
communities. Numerous algorithms have been proposed to address the
problem of classifying nodes into communities. Interestingly,
different algorithms employ highly diverse methods and
transformations, or measures, of the network topology. For example,
many algorithms maximize the modularity index \cite{community5} by a
wide spectrum of optimization techniques
\cite{community7,community6,community9,community11}. Others exploit
quantities such as betweeness centrality
\cite{community1,community5}, traces of random walk
\cite{community1,community2,community5,community13}, eigenvectors of
the network Laplacian \cite{community6,community10,community11},
electrical conductance \cite{community3}, and others. While some
algorithms recursively split the network into communities separated
by ``weak links''
\cite{community1,community2,community3,community7,community9,community11},
others take the bottom-top approach and recursively merge highly
similar communities, based on various similarity indices
\cite{community5,community6}. Also, while many algorithms output a
dendogram (a tree) with partition of the network into disjoint
communities at all possible levels of resolution, other studies
provide an overlapping community structure; for example, based on
identification of almost complete subgraphs \cite{community8} or
mapping to magnetic domains \cite{community4}.

\begin{figure}
\vspace{-1cm} \includegraphics[width=6cm]{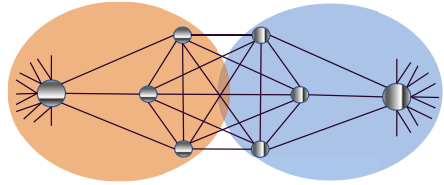} \vspace{-1.5cm}
\caption{Illustration of the difference between basins of attraction
and network communities. In the plotted toy network, the six nodes
in the middle are fully connected and clearly form a single
community. However, the three nodes on the left side are attracted
to the left hub, whereas the three nodes on the right side are
attracted to the right hub. Thus, they are split between two
different basins of attraction (indicated with different background
colors and different orientation of node fill patterns).}
\label{comm_fig}
\end{figure}

How is the partition into basins of attraction compared to other
community detectors? First, most algorithms are \emph{global}, since
they utilize as much information as possible about the network
topology to improve the identification of the communities. In
contrast, few other methods (e.g.,
\cite{community8,comm_local1,comm_local2,comm_local3}), including
our basins of attraction, are computed in a \emph{local} manner---
each node is assigned to a community based only on its immediate
neighborhood. Second, and more important, the goal of most community
detectors is to find a partition that maximizes intra-community
proximity and inter-community separation. That usually takes the
form of maximizing the number of links within a community while
minimizing the number of links between communities. As opposed to
that, our partition to basins of attraction addresses a different
question: which nodes are affiliated with the same hub? While in
many cases this attribute is correlated with community structure,
this is not necessarily always the case, as we demonstrate in Figure
\ref{comm_fig}.

A possible outcome of our analysis is revealed when we test two
real-life networks for which the problem of basins is of practical
importance: The Internet at the Autonomous Systems (AS) level
\cite{DIMES} as of 2007, and the energy landscape's network of
Lennard-Jones clusters \cite{Doye}. Both networks are scale-free,
with $\gamma=2.5,2.9$, respectively. In both networks there is a
giant basin which attracts most nodes (with Verizon's AS being the
peak in the Internet), and a few tiny basins, in agreement with the
theoretical results for the model scale-free networks. In the
Lennard-Jones network, an uphill walk in the degree space, which can
be mapped in general onto a downhill walk in the energy landscape,
will end up at the node of highest degree, which can be interpreted
as the ideal glass state \cite{SupercooledReview}. Note the
different situation for the energy landscape of proteins, where the
energy increases with the degree, such that the system is expected
to follow a downhill walk in the degree space
\cite{gradient_protein}.

For the Internet, the existence of a giant basin implies that a
routing scheme that forwards all messages in a steepest ascent
manner will quickly arrive at the hub. From the hub, messages could
be routed to their target according to a predefined target-specific
sequence embedded in the packet, as was previously suggested
\cite{KrioukovFall,routing}. This leads to an efficient routing
scheme which requires practically no knowledge of the network
topology at the nodes, and is thus highly scalable. An obvious
drawback of such scheme is the congestion generated at the hubs,
which is eliminated in other methods (for example, by routing
through shortest paths when the hubs are avoided \cite{bottleneck},
or by walking down the congestion gradient \cite{gradient_routing})
Therefore, the steepest ascent search might not be of immediate
applicability to the Internet itself, but is however of interest in
other newly designed communication networks where the hubs can carry
high load. In this context, we note the interesting fact that
Bogu\~na {\it et. al.} \cite{Navigation} also find a transition
between navigable and non-navigable network topology at $\gamma
\approx 2.6$, although in their case the navigation is based on
minimizing distances within a hidden metric space.

Our partitioning has another potential practical applications for
locating the node of highest degree in various search scenarios. A
local search starting from a single node and following a steepest
ascent would always be successful in networks with a single basin of
attraction, as in scale-free networks with $\gamma<\gamma_c$. With
more than one basin, a strategy could be devised for starting from a
number of randomly selected nodes to find the highest degree with a
prescribed rate of success.

A concrete example for such an application is routing in wireless
sensor networks \cite{Sensor1}. A wireless sensor network is a
system consisting of spatially distributed autonomous devices using
sensors to cooperatively monitor physical or environmental
conditions. In a typical sensor network, one distinguished node
serves as a gateway between the sensors and the end users, and must
collect data from the nodes. Since energy is usually a very scarce
resource at the nodes, an efficient protocol must be designed to
transmit the measured data to the base station. Thus, our steepest
ascent protocol, in which each node sends out data to its neighbor
with highest degree, is of interest. This protocol is expected to be
relevant in heterogeneous sensor networks, in which the
communication range varies between the nodes \cite{Sensor2}. Indeed,
we found (data not shown) that for a power-law distribution of
communication ranges $\Phi(r)\sim r^{-\epsilon}$, there exist a
regime in ($\epsilon$,$\av{r}$) parameter space for which the
network collapses into a single basin, making the steepest ascent
protocol highly efficient.


\begin{acknowledgments}
We thank E.~M.~Bollt, G.~Oshanin and D.~Krioukov for discussions, and
H.~D.~Rozenfeld for discussions and for supplying an
hierarchical networks generator and help with the graphics.
Financial support from ONR, NSF, the Israel Science
Foundation, and the Israel Center for Complexity Science, is
gratefully acknowledged. S.C. is supported by the Adams Fellowship
Program of the Israel Academy of Sciences and Humanities.
\end{acknowledgments}

\appendix
\section{Alternative Definition of Basins}
\label{alternative}

When the weights of the nodes are taken from a discrete
distribution, as in the case where the weight is the degree of the
node, neighboring nodes may have the same weight.  A method is then
required to break the tie.  In Section \ref{DefSect} we presented an
algorithm that overcomes this difficulty, which we term the
\emph{local search} algorithm. The following  \emph{recursive
search} algorithm works as well.

Suppose the search starts at node $i$, and let $j_\textrm{max}$ be
the neighbor(s) of $i$ of highest degree $k_{j_\textrm{max}}$.
Denote the number of neighbors with degree $k_{j_\textrm{max}}$ as
$q$.

\begin{enumerate}
\item If $k_i>k_{j_\textrm{max}}$, $i$ is a peak.
\item If $k_i<k_{j_\textrm{max}}$, $i$ is attracted to
$j_\textrm{max}$. (If there is more than one neighbor with degree
$k_{j_\textrm{max}}$ (i.e., $q>1$), select one randomly.)
\item If $k_i=k_{j_\textrm{max}}$, mark $i$ as \emph{visited} and look for
the attractor of $j_\textrm{max}$, recursively, among unvisited
nodes. If $q>1$, look also for the attractors of all other neighbors
of $i$ with degree $k_{j_\textrm{max}}$. Keep only the attractor of
highest degree among the $q$ attractors.
\item If the degree of the attractor of $j_\textrm{max}$ is larger than $k_i$,
$i$ is attracted to $j_\textrm{max}$. If the degree of the attractor
of $j_\textrm{max}$ equals $k_i$, $i$ is a peak.
\end{enumerate}
In other words, in a search for a peak strictly higher than its
neighbors, we are allowed to surf over ``ridges'' of connected nodes
of equal degree, until either reaching a peak or a dead end.

\begin{figure}
\includegraphics[width=6cm]{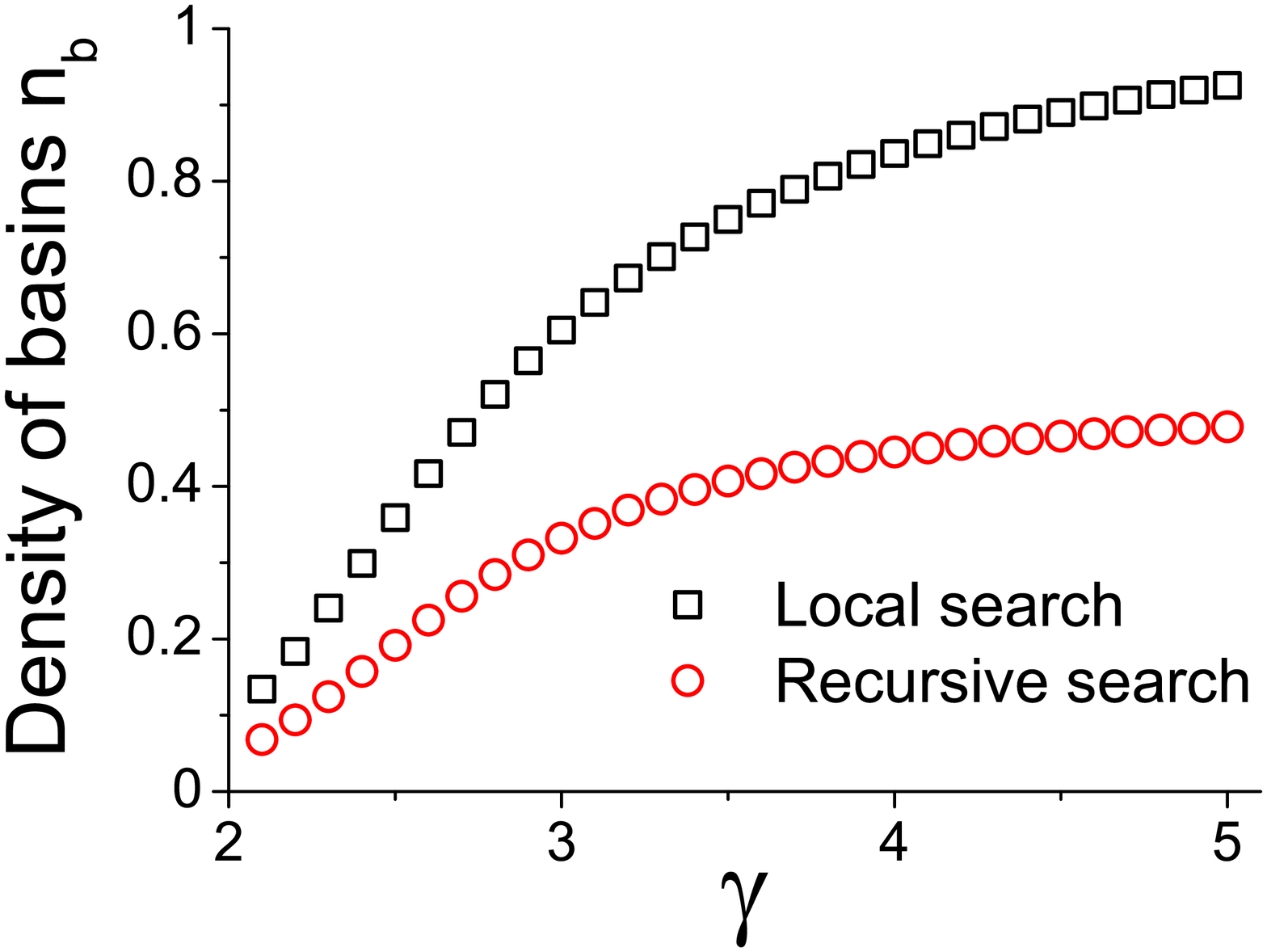}
\includegraphics[width=6cm]{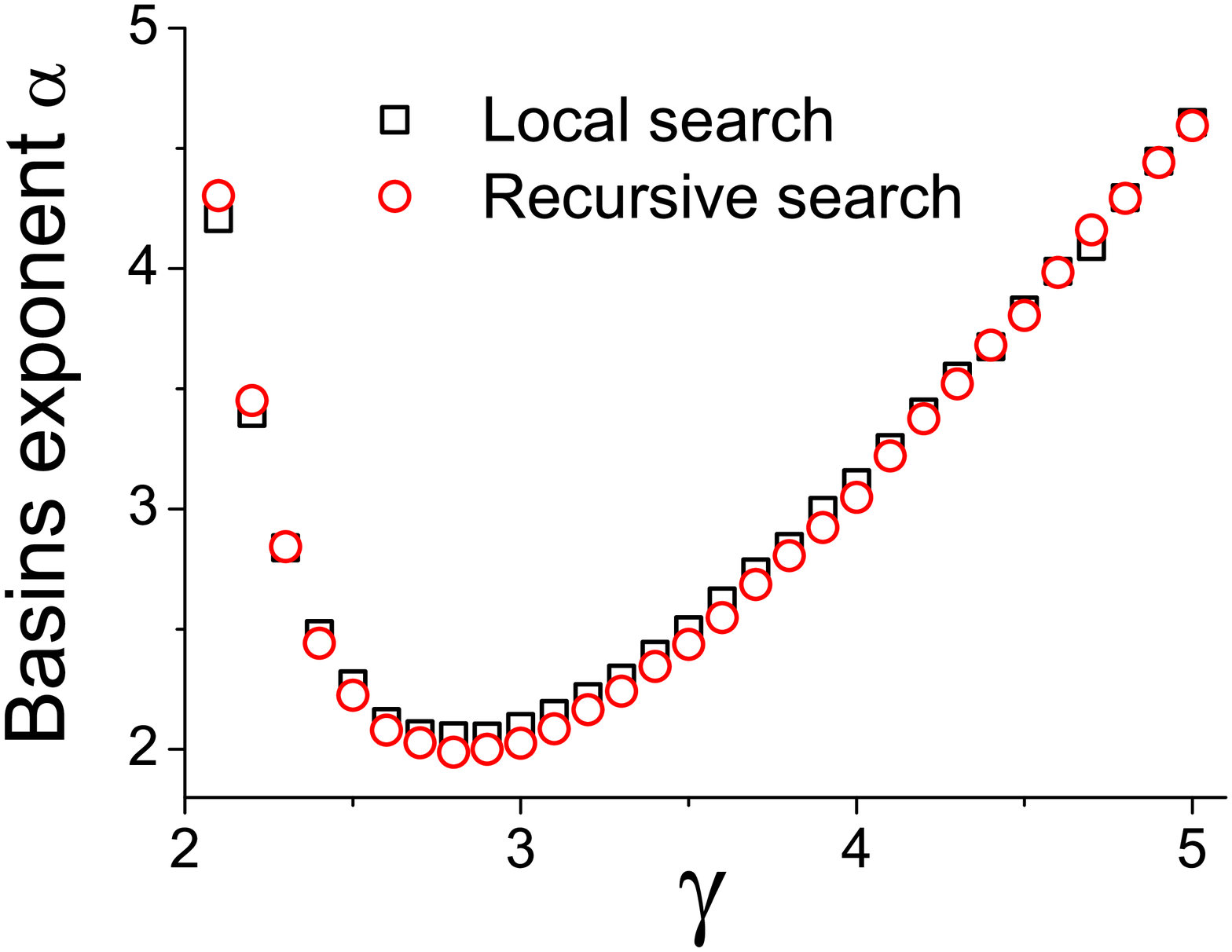}
\caption{Basin topology with local vs.~recursive search. (a)~Density
of basins, and (b)~the basin exponent $\alpha$ as a function of
$\gamma$, in SF networks with $N=1000$ and $m=1$.} \label{DFS}
\end{figure}

Despite the broad distribution of degrees in SF networks, the
majority of the nodes have the minimal degree $m$, or a degree close
to $m$. Thus, one may expect many ridges to form and as a result, a
different basin count, depending on whether the local or recursive
search is employed. For example, in the hierarchical networks
studied in Section \ref{flowers}, a recursive search yields a single
giant basin for \emph{all} $(1,v)$-nets. In random SF networks with
large $\gamma$ the recursive search method also yields fewer basins
(Fig.~\ref{DFS}(a)), which is explained by the prevalence of ridges,
in this case, due to the high density of small-degree nodes.
However, broader properties of the basins topology remain unaffected
by the search algorithm: $Q(s)$ is practically the same, for large
$s$, as is also the basin exponent $\alpha$, extracted from either
method (Fig.~\ref{DFS}(b)).

\section{Properties of the Number of Basins in SF Networks}
\label{theory_app}

In this appendix, we calculate two quantities related to the number
of basins.

\subsection{Variance of the number of basins}

\begin{figure}
\includegraphics[width=8cm]{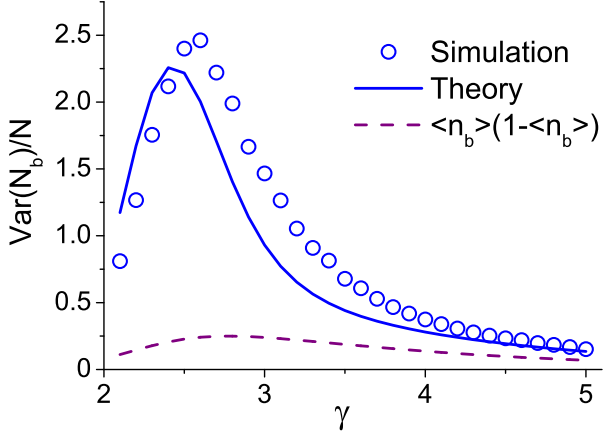}
\caption{Variance in number of basins  for SF networks with $N=1000$
and $m=1$. Symbols represent simulation results, and the solid line
corresponds to Eq.~(\ref{var_2neigh}). The dashed line is obtained
by neglecting the covariance term and has the form
$\av{n_b}(1-\av{n_b})$ (where $\av{n_b}$ is calculated from
Eq.~(\ref{p_peak})).} \label{Variance_fig}
\end{figure}

Denote by $A_i$ the indicator of the event that node $i$ is a peak,
such that $N_b=\sum_{i=1}^{N}A_i$. To compute the variance, we shall use the general formula
\begin{equation}
\mbox{Var}(N_b)=\sum_{i=1}^{N}\mbox{Var}(A_i)+2\sum_{i=1}^{N}\sum_{j>i}^{N}\mbox{Cov}(A_i,A_j)
\end{equation}
The first term on the right-hand side is easy to compute:
\begin{equation*}
\sum_{i=1}^{N}\mbox{Var}(A_i)=\sum_{i=1}^{N}P\{A_i\}(1-P\{A_i\})=N\av{n_b}(1-\av{n_b})
\end{equation*}
Here $P\{A_i\}$ is the probability for $A_i$ to occur. For the
second term, we get
\begin{eqnarray*}
\mbox{Cov}(A_i,A_j)&=&\av{A_iA_j}-\av{A_i}\av{A_j} \\
&=&
P\{A_iA_j\}-P\{A_i\}P\{A_j\}
\end{eqnarray*}
Let nodes $i$ and $j$ have degrees $k_1$ and $k_2$, respectively.
What is the probability for both nodes $i$ and $j$ to be peaks? We
condition this probability on whether $i$ and $j$ are connected,
which is $k_1k_2/(N\av{k})$ \cite{agata}. If they are connected, and
they have different degrees, clearly only one of them can serve as a
peak, so only the case when both have the same degree $k$
contributes to the covariance. Also, we have to take into account
that $i$ and $j$ might share common neighbors. Thus, the probability
of $i$ to be a peak is enhanced if $j$ is known to be one. We make
the approximation that the number of common neighbors $c$ is fixed
once $k_i$ and $k_j$ are given, and is given by:
\begin{equation*}
c_{k_i,k_j}\approx (N-2)\sum_{k_{\ell}=m}^{\infty}{\cal
P}(k_{\ell})\frac{k_ik_{\ell}}{N\av{k}}\frac{k_jk_{\ell}}{N\av{k}}\approx
\frac{k_ik_j\av{k^2}}{N\av{k}^2},
\end{equation*}
since this is the probability, summing over all possible degrees of
the node $\ell \ne i,j$, that it is adjacent to both $i$ and $j$.
For both $i$ and $j$ to be peaks, if $k_i<k_j$, $k_i$ nodes need to
have degree less than $k_i$, but only $k_j-c$ nodes need to have
degree less than $k_j$ (since the $c$ common nodes are guaranteed to
have degree less than $k_i<k_j$), and vice-versa if $k_i>k_j$.
Approximating the probabilities for two nodes without common
neighbors to be peaks as independent, we get,
\begin{eqnarray}
\label{var_2neigh}
&&\mbox{Cov}(A_i,A_j)=\sum_{k=m}^{\infty}k^2[{\cal
P}(k)]^2[f(k)]^{2(k-1)-c_{k,k}}/(N\av{k}) \nonumber \\ &+&
2\sum_{k_1=m}^{\infty}\sum_{k_2>k_1}{\cal P}(k_1){\cal P}(k_2)
\left(1-\frac{k_1k_2}{N\av{k}}\right)
[f(k_1)]^{k_1}[f(k_2)]^{k_2-c_{k_1,k_2}} \nonumber \\ &+&
\sum_{k=m}^{\infty}[{\cal P}(k)]^2
\left(1-\frac{k^2}{N\av{k}}\right)[f(k)]^{2k-c_{k,k}} \nonumber
\\ &-&
\left[\sum_{k=m}^{\infty}{\cal P}(k)[f(k)]^{k}\right]^2,
\end{eqnarray}
where $f(k) = \sum_{k'=m}^{k}\frac{k'{\cal P}(k')}{\av{k}}$ is the
probability for a neighbor to have degree no larger than $k$. The
first term corresponds to the case where the nodes are directly
connected and have identical degree; the second term is the case
when they are not directly connected, and have different degrees; in
the third term they are not directly connected but have equal
degree; and the last term is just $P\{A_i\}P\{A_j\}=P\{A_i\}^2$.
This formula is compared to simulations in Fig.~\ref{Variance_fig}
to find a qualitative agreement. We also plot the
$\av{n_b}(1-\av{n_b})$ term alone, neglecting the covariance, and
find that it is a good approximation for the case of large $\gamma$.

\subsection{Density of basins of size one}

Consider a given node $i$ with degree $k$ and take one of its
neighbors $j$; suppose this neighbor $j$ has degree $k'$. For $i$ to
be a peak, $k'$ must be less than or equal to $k$. For $i$ to form a
basin of size one, $j$ must have at least one neighbor of degree
$k+1$ or above, in order to be attracted to that neighbor and not to
$i$. If we assume that at least up to the second shell, $i$ is a
root of a tree, we have:
\begin{equation*}
\label{q1_eq}
Q(1) = \sum_{k=m}^{\infty}{\cal
P}(k)\left[\frac{k{\cal
P}(k)}{\av{k}}+\sum_{k'=m}^{k-1}\frac{k'{\cal
P}(k')}{\av{k}}q(k,k')\right]^k,
\end{equation*}
where
\begin{equation*}
q(k,k') \equiv 1-\left[\sum_{k''=m}^{k}\frac{k''{\cal
P}(k'')}{\av{k}}\right]^{k'-1}
\end{equation*}
is the probability that at least one of the $k'-1$ neighbors (others
than $i$) of $j$ has degree above $k$. We wrote a separate term for
the case of $k'=k$, since in this case we are guaranteed that $j$ is
not attracted to $i$, regardless of the degrees of the neighbors of
$j$. The small correction due to the case when $j$ has another
neighbor (other than $i$) of degree exactly $k$ can be calculated
analytically as well, but was found to be negligible. For $\gamma
\rightarrow 2$, when $\av{k} \sim \log{N}$, only the $k=m$ term is
significant, and thus $Q(1) \sim {\cal P}(m)\left[m{\cal
P}(m)\right]^m (\ln N)^{-m}$, and almost all basins are solitary
(see Eq. (\ref{gamma2})). This is confirmed in the simulations (Fig.
\ref{NumBasins}).

\section{Basins of Attraction in One Dimension}
\label{der}

We look at the distribution of basins of attraction in
one-dimensional lattices. Consider first a valley separated by
distance $i$ from the peak on the left and distance $j$ from the
peak on the right, such that particles from both peaks belong to its
basin of attraction. The probability of this is
\begin{equation}
\label{R++} R^{++}_{ij}=\int_0^1dh\, \Pi^+_{i}(h)\Pi^+_j(h)\,,
\end{equation}
where, e.g., $\Pi^+_j(h)$ is the probability that $j$ heights to the
right of the valley of height $h$ are ascending and the last height
is the peak which belongs to the basin of attraction of our valley.
The probability $\Pi^+_k(h)$ admits an integral representation
\begin{equation*}
\Pi^+_k(h)=\int_{\substack{h<x_1<\ldots<x_k<1\\x_{k-1}<x_{k+1}<x_k}}
\prod_{a=1}^{k+1} dx_a\,.
\end{equation*}
Integrating over $x_1,\ldots,x_{k-1}$ we recast the above integral
into
\begin{equation*}
\Pi^+_k(h) = \int_{h<x_{k+1}<x_k<1}
dx_k\,dx_{k+1}\,\frac{(x_{k+1}-h)^{k-1}}{(k-1)!}\,,
\end{equation*}
and the remaining integration is trivial:
\begin{equation}
\label{Pi+} \Pi^+_k(h) = \frac{(1-h)^{k+1}}{(k+1)!}\,.
\end{equation}
Inserting this equation into \eqref{R++} we obtain
\begin{equation}
\label{Rij++} R^{++}_{ij}=\frac{1}{(i+1)!(j+1)!}\,\frac{1}{i+j+3}\,.
\end{equation}

Similarly, we compute
\begin{equation}
\label{R+-} R^{+-}_{ij}=\int_0^1dh\, \Pi^+_{i}(h)\Pi^-_j(h)\,,
\end{equation}
where $\Pi^-_j(h)$ is the probability that $j+1$ heights to the
right of the valley of height $h$ are ascending and the last height
is the peak which belongs to the basin of attraction of the next
valley (to its right). The probability $\Pi^-_k(h)$ can be written
as
\begin{equation*}
\Pi^-_k(h)=\int_{\substack{h<x_1<\ldots<x_{k+1}<1\\x_{k+1}>x_{k+2}<x_k}}
\prod_{a=1}^{k+2} dx_a\,.
\end{equation*}
The two last integrations are easily performed,
\begin{equation*}
\Pi^-_k(h)=\int_{h<x_1<\ldots<x_{k}<1} x_k(1-x_k) \prod_{a=1}^{k}
dx_a\,.
\end{equation*}
Integrating over $x_1,\ldots,x_{k-1}$ we recast the above integral
into
\begin{equation*}
\Pi^-_k(h)=\int_h^1 dx_k\, x_k(1-x_k)\,\frac{(x_k-h)^{k-1}}{(k-1)!}\,,
\end{equation*}
which is then computed to yield
\begin{equation}
\label{Pi-} \Pi^-_k(h) =
\frac{(1-h)^{k+1}}{(k+1)!}\left[1-(1-h)\,\frac{2}{k+2}\right]\,.
\end{equation}
Plugging \eqref{Pi+} and \eqref{Pi-} into \eqref{R+-} we obtain
\begin{eqnarray}
\label{Rij+-}
R^{+-}_{ij}&=&\frac{1}{(i+1)!(j+1)!}\,\frac{1}{i+j+3}\nonumber\\
&-& \frac{1}{(i+1)!(j+2)!}\,\frac{2}{i+j+4}\,.
\end{eqnarray}

Since $R^{-+}_{ij}=R^{+-}_{ji}$, the last quantity to compute is
\begin{equation}
\label{R--} R^{--}_{ij}=\int_0^1dh\, \Pi^-_{i}(h)\Pi^-_j(h)\,.
\end{equation}
Using \eqref{Pi-} we get
\begin{eqnarray}
\label{Rij--}
R^{--}_{ij}&=&\frac{1}{(i+1)!(j+1)!}\,\frac{1}{i+j+3}\nonumber\\
&-& \frac{1}{(i+1)!(j+1)!}\,\frac{2}{i+j+4}\left[\frac{1}{i+2}+\frac{1}{j+2}\right]\nonumber\\
&+& \frac{1}{(i+2)!(j+2)!}\,\frac{4}{i+j+5}\,.
\end{eqnarray}

Equation \eqref{Rij++} is valid when $i\geq 1, j\geq 1$, and the
size of the basin of attraction is $s=i+j+1\geq 3$. Overall, the
density of basins of attraction of type $++$ of size $s$ is
\begin{equation}
\label{Rs++} R^{++}(s) = \sum_{\substack{i\geq 1, j\geq
1\\i+j=s-1}}R^{++}_{ij}=\frac{2^{s+1}-2}{(s+2)!}- \frac{2}{(s+2)s!}\,.
\end{equation}

Equation \eqref{Rij+-} is valid when $i\geq 1, j\geq 0$. The density
of basins of attraction of type $+-$ of size $s$ is
\begin{equation*}
R^{+-}(s) = \sum_{\substack{i\geq 1, j\geq 0\\i+j=s-1}}R^{+-}_{ij}\,.
\end{equation*}
Computing the sum we find
\begin{eqnarray}
\label{Rs+-}
R^{+-}(s)&=& \frac{2^{s+1}-2}{(s+2)!}- \frac{1}{(s+2)s!} \nonumber\\
&-& 2\,\frac{2^{s+2}-2}{(s+3)!} + \frac{4}{(s+3)(s+1)!}\,.
\end{eqnarray}
Of course, $R^{+-}(s)= R^{-+}(s)$, and \eqref{Rs+-} is valid when
$s\geq 2$.

Equation \eqref{Rij--} is valid when $i\geq 0, j\geq 0$. The density
of basins of attraction of size $s$ and type $--$ is given by
\begin{equation*}
R^{--}(s) = \sum_{\substack{i\geq 0, j\geq 0\\i+j=s-1}}R^{--}_{ij}\,.
\end{equation*}
Computing the sum we find
\begin{eqnarray}
\label{Rs--} R^{--}(s)&=& \frac{2^{s+1}-2}{(s+2)!}-
4\,\frac{2^{s+2}-2}{(s+3)!}
+ \frac{4}{(s+3)(s+1)!}\nonumber\\
&+& 4\,\frac{2^{s+3}-2}{(s+4)!} -  \frac{8}{(s+4)(s+2)!},
\end{eqnarray}
which is valid for $s \geq 1$. Defining $R^{++}(1)=R^{++}(2)=0$ and
$R^{+-}(1)=R^{-+}(1)=0$, we finally have ($s \geq 1$):
\begin{equation}
\label{Rs_all} R(s) = R^{++}(s) + R^{+-}(s) + R^{-+}(s) + R^{--}(s)\,.
\end{equation}
Inserting
\eqref{Rs++},\eqref{Rs+-}, and \eqref{Rs--} into \eqref{Rs_all} we
arrive at the announced result  \eqref{Rs}.

\bibliography{ckb}

\end{document}